\documentclass{raa}            

\usepackage{graphicx,times}             

\begin{document}

   \title{GSC 4560--02157: a New Long-period Eclipsing
Cataclysmic Variable Star}

   \volnopage{Vol.0 (200x) No.0, 000--000}      
   \setcounter{page}{1}          

   \author{A. V. Khruslov
      \inst{1}
      \and
 A.V. Kusakin\inst{2}
 \and
 E.A. Barsukova\inst{3}
 \and
 V.P. Goranskij\inst{4}
 \and
 A.F. Valeev\inst{3}
 \and
 N.N.~Samus\inst{1,4}}


   \institute{Institute of Astronomy, Russian Academy of Sciences,
48, Pyatnitskaya Str., Moscow 119017, Russia; {\it khruslov@bk.ru}\\
        \and
             Fesenkov Astrophysical Institute,
Almaty 050020, Kazakhstan\\
        \and
             Special Astrophysical Observatory,
Russian Academy of Sciences, Nizhny Arkhyz, Karachai-Cherkessian
Republic 369167, Russia\\
        \and
             Sternberg Astronomical Institute,
Lomonosov Moscow State University, 13 University Ave., Moscow
119092, Russia\\
   }

   \date{Received~~2014 month day; accepted~~2014~~month day}

\abstract{We study the newly discovered variable star GSC
4560--02157. CCD photometry was performed in 2013--2014, and a
spectrum was obtained with the 6-m telescope in June, 2014. GSC
4560--02157 is demonstrated to be a short-period ($P=0.265359^d$)
eclipsing variable star. All its flat-bottom primary minima are
approximately at the same brightness level, while the star's
out-of-eclipse brightness and brightness at secondary minimum
varies considerably (by up to $0.6^m$) from cycle to cycle.
Besides, there are short-term (time scale of 0.03--0.04 days)
small-amplitude brightness variations out of eclipse. This
behavior suggests cataclysmic nature of the star, confirmed with a
spectrum taken on June 5, 2014. The spectrum shows numerous
emissions of the hydrogen Balmer series, HeI, HeII.
 }

   \authorrunning{A. V. Khruslov et al.}            
   \titlerunning{GSC 4560--02157, a New Eclipsing
Cataclysmic Variable Star}  

   \maketitle

%
%
\section{Introduction}           
\label{sect:intro}

Among short-period spectroscopic and eclipsing binaries,
cataclysmic stars are of special interest. These systems contain a
white dwarf and an ordinary dwarf, yellow or red, which donors
matter to the Roche lobe of the white dwarf. Matter flows,
accretion disks, hot spots (or hot lines) at the interaction
between the flow and the disk, all cause complicated light
variability, e.g., changing brightness levels at different
activity states, outbursts, quasi-periodic variations, humps on
the light curve, etc. For a general review of cataclysmic variable
stars, see Warner~\cite{cv}.

In this paper, we present the discovery and study of the new
variable star GSC 4560--02157. It is shown to be a short-period
eclipsing variable star exhibiting, outside its primary minima,
strong variations of the brightness level, with superimposed
short-term brightness variations. The cataclysmic nature of the
new variable could be confirmed with its spectrum taken with the
6-meter telescope of the Special Astrophysical Observatory.

\section{Discovery and photometric observations}

Variability of GSC 4560--02157 was suspected by one of
the authors (A.V.~Khruslov) in 2005 from the publicly available
data of the Northern Sky Variability Survey (NSVS;
http://skydot.lanl.gov/nsvs/nsvs.php, Wozniak et al. \cite{nsvs}).
The star is in $14''$ from GSC 4560--02269, and NSVS data refer to
a blend of the two stars. In March 2013 (JD 2456364), we started
CCD photometry of the two stars in order to establish which of
them varies.

The actual variable star is GSC 4560--02157 ($15^h43^m36.65^s$,
$+75^\circ15'41.1''$, J2000.0 in the 2MASS Point Source Catalog,
Skrutskie et al. \cite{2mass}).

Our CCD observations in the Johnson $R$ and $V$ bands were
performed at the Tien Shan Astronomical Observatory of the
V.G.~Fesenkov Astrophysical Institute, at the altitude of 2750~m
above the sea level. The observatory has two Zeiss 1000-mm
telescopes. Most of our observations were performed with the
eastern Zeiss 1000-mm reflector (the focal length of the system
was 6650~mm, the detector being an Apogee U9000 D9 CCD camera).
During the three last nights (JD 2456772--2456784), we used the
newly introduced western Zeiss 1000-mm reflector (the focal length
of the system was 13250~mm, the detector being an Apogee F16M CCD
camera). In the $R$ band, we obtained 2455 brightness measurements
(JD 2456364--2456784), and in the $V$ band, 828 measurements (JD
2456739--2456784). We performed reductions using the MaxIm DL
aperture photometry package.

The total range of the star's brightness variations is between
14.40$^m$ and 15.53$^m$ $R$, 14.75$^m$ and 16.10$^m$~$V$.

Table~1, available in its complete form in the electronic
supplement to this paper, contains our photometric observations.
Here we present its first lines (photometry in each of the colors)
for guidance concerning its format.

\begin{table}
\caption{Observations} \label{table:1}
\centering
   \begin{tabular}{ccc|ccc|ccc}
   \hline
HJD & $V$, mag & Err &HJD & $V$, mag & Err &HJD & $V$, mag & Err\\
\hline
2456740.1422 & 15.726 & 0.017& 2456740.1967 & 15.439 & 0.021& 2456740.2529 & 15.586 & 0.013\\
2456740.1436 & 15.785 & 0.017& 2456740.1981 & 15.448 & 0.016& 2456740.2543 & 15.578 & 0.014\\
2456740.1450 & 15.860 & 0.017& 2456740.2010 & 15.424 & 0.016& 2456740.2556 & 15.616 & 0.015\\
2456740.1464 & 15.848 & 0.017& 2456740.2023 & 15.418 & 0.015& 2456740.2570 & 15.658 & 0.012\\
2456740.1478 & 15.791 & 0.020& 2456740.2037 & 15.432 & 0.017& 2456740.2599 & 15.766 & 0.015\\
...\\
\hline
HJD & $R$, mag & Err &HJD & $R$, mag & Err &HJD & $R$, mag & Err\\
\hline
2456364.4342 & 14.974 & 0.005 & 2456371.4766 & 15.065 & 0.011 & 2456374.3476 & 15.463 & 0.006\\
2456364.4353 & 14.971 & 0.005 & 2456371.4777 & 15.054 & 0.011 & 2456374.3488 & 15.453 & 0.011\\
2456364.4365 & 15.006 & 0.005 & 2456371.4789 & 15.043 & 0.009 & 2456374.3500 & 15.455 & 0.010\\
2456364.4376 & 14.996 & 0.006 & 2456371.4801 & 15.036 & 0.009 & 2456374.3512 & 15.433 & 0.012\\
2456364.4387 & 15.002 & 0.011 & 2456371.4813 & 15.024 & 0.006 & 2456374.3524 & 15.450 & 0.010\\
...\\
\hline \hline
\end{tabular}
\end{table}

The finding chart (Fig. 1) identifies the variable star,
comparison star, and check star. The comparison star was GSC
4560--01221 ($15^h43^m36.53^s$, $+75^\circ18'04.2''$, J2000.0),
and the check star, GSC 4560--01352 ($15^h43^m03.53^s$,
$+75^\circ17'55.1''$, J2000.0). The magnitudes of the comparison
star in the GSC2.3 catalog (Lasker et al. \cite{g23}) are $R =
14.11$ and $V = 14.33$.

\begin{figure}
\centering
\includegraphics[width=12cm]{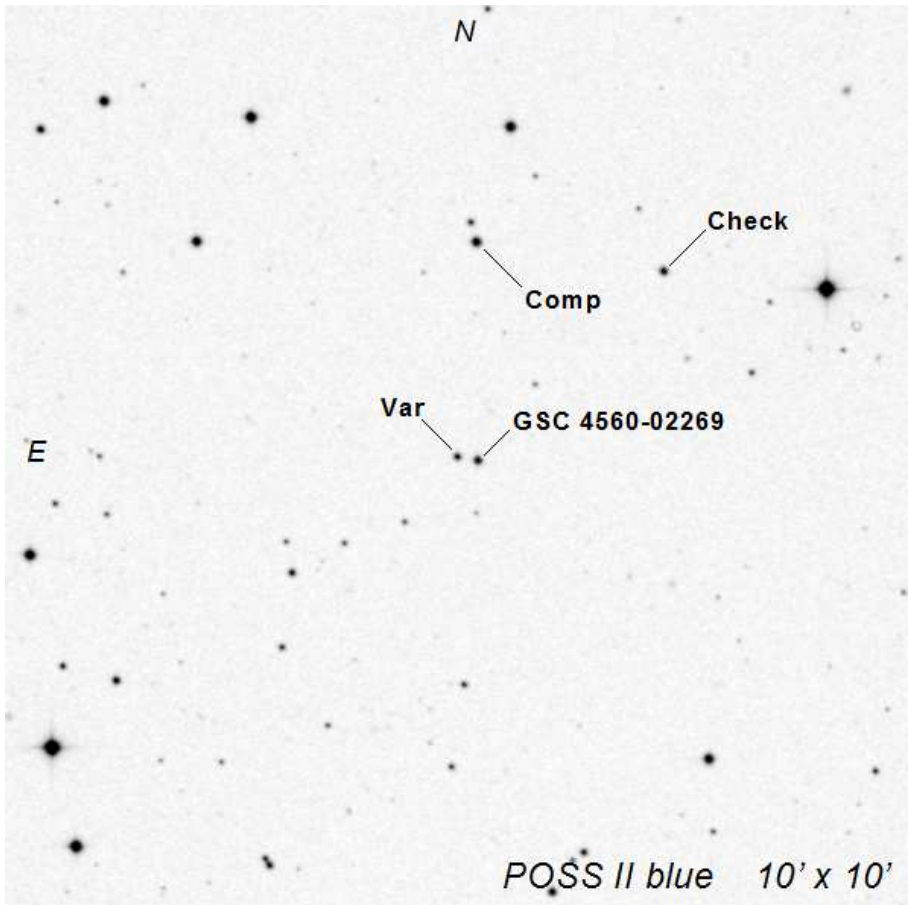}
\caption{The finding chart, with the variable, its close neighbor,
comparison star, and check star marked} \label{fig:1}
\end{figure}

\section{Results}

We analyzed the time series using the Lafler--Kinman method
(Lafler \& Kinman \cite{lafkin}) implemented in the WinEfk code
developed by one of the authors (V.P.G.). GSC 4560--02157 is found
to be an Algol-type eclipsing binary with a short period (about
0.265 days) and considerable brightness  variations outside the
primary eclipse, which remains at approximately the same magnitude
for all the observed cycles. The star's brightness outside the
primary eclipse slowly changes its level between different orbital
cycles by as much as $0.5^m{-}0.6^m$; small short-term brightness
variations are superimposed. In any individual cycle, the star is
redder in the primary minimum than outside it. The ephemeris for
the primary eclipses is:
$${\rm HJD(min)} = 2456719.314 + 0.265359^d\times E.$$ The light
curve folded with the orbital period is displayed in Fig.~2. The
duration of the primary minimum is $D = 0.18 P = 0.05^d$, and the
eclipse is apparently total ($d = 0.05 P = 0.012^d$). The level of
the secondary minimum varies with general changes of the
brightness level outside the primary eclipse. The light curve
folded with the orbital period is displayed in Fig.~2.

\begin{figure}
\centering
\includegraphics[width=12cm]{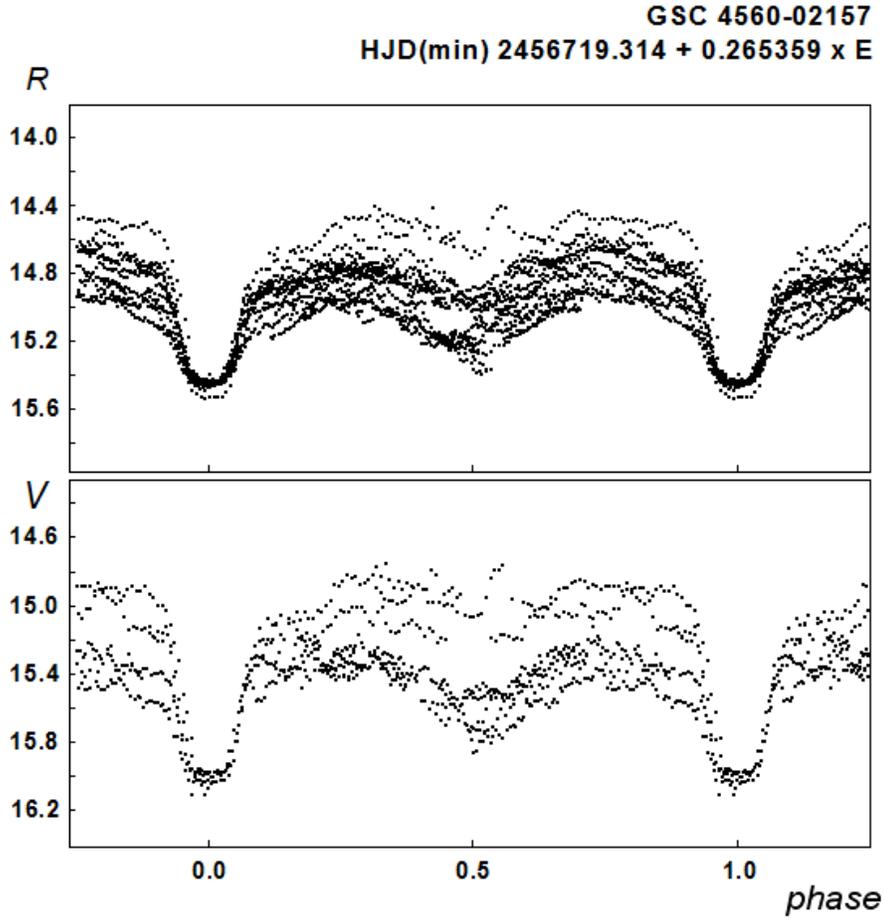}
\caption{The
light curves of GSC 4560--02157 in the $R$ and $V$ bands, folded
with the orbital period}
\label{fig:2}
\end{figure}

For comparison, Fig.~3 shows the light curve of the known dwarf
nova GY~Cnc, folded with its orbital period, $0.1754424988^d$
(Feline et al. \cite{feline}). The observations we use are those
available from the Catalina Surveys (Drake et al. \cite{catal}).
The scattered bright data points correspond to dwarf-nova
outbursts. The period of the binary GY~Cnc is somewhat shorter
than that of GSC 4560--02157, and no outbursts of GSC 4560--02157
have been detected so far, but the general character of the
eclipsing light curves is very similar for the two stars. The
spectrum of GY~Cnc reproduced in Szkody et al. \cite{szkody}
resembles that of GSC 4560--02157 very strongly.

\begin{figure}
\centering
\includegraphics[width=12cm]{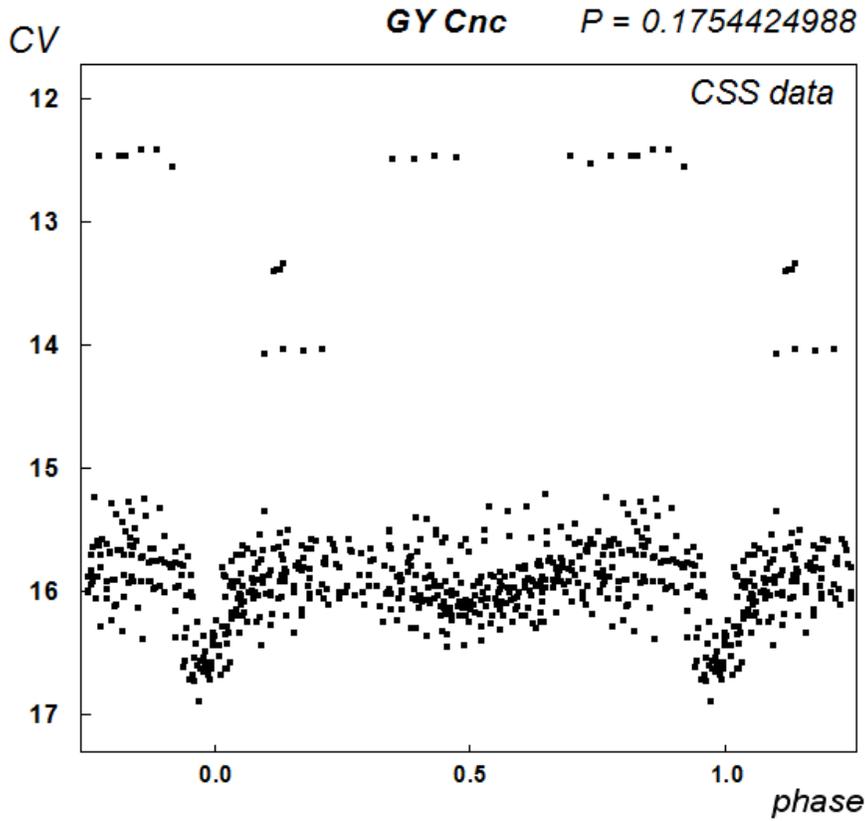}
\caption{The light
curve of the known dwarf nova GY Cnc, based on Catalina Surveys
data and folded with the orbital period. The data points above the
eclipsing light curve correspond to outbursts} \label{fig:3}
\end{figure}

Individual cycles outside the primary eclipse reveal
small-amplitude short-period brightness variations (Fig.~4). To
study them, we analyzed the more complete 2014 $R$-band data.
Observations during primary eclipses were excluded, and the
remaining data were reduced to the same brightness level, thus
eliminating comparatively slow outside-eclipse brightness level
variations. The total of 1637 brightness estimates obtained in
this way (JD 2456716--2456784) were subject to frequency analysis
using Deeming's method (Deeming \cite{deem}). A part of the power
spectrum is shown in Fig.~5. Table~2 lists the six most prominent
frequencies. These variations had the largest amplitude (up to
$0.3^m$) on the last night, JD 2456784.

\begin{table}
\caption{Frequencies Detected Outside Eclipses} \label{table:2}
\centering
\begin{tabular}{cccc}
\hline \hline
ID & $F$, c/d & $P$, days &Semi-amplitude,\\
&&& mag\\
\hline
$f_1$&24.8432&0.0402525&0.0171\\
$f_2$&28.4026&0.0352080&0.0165\\
$f_3$&30.5982&0.0326817&0.0138\\
$f_4$&31.9354&0.0313132&0.0120\\
$f_5$&27.7483&0.0360383&0.0114\\
$f_6$&20.7149&0.0482744&0.0100\\
\hline
\end{tabular}
\end{table}

\begin{figure}
\centering
\includegraphics[width=12cm]{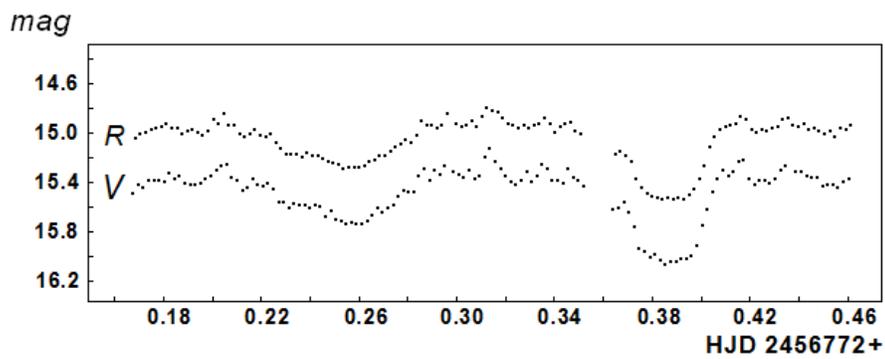}
\caption{The
light curves of GSC 4560--02157 in the $R$ and $V$ bands from a single orbital cycle,
showing short-period brightness variations} \label{fig:4}
\end{figure}

\begin{figure}
\centering
\includegraphics[width=12cm]{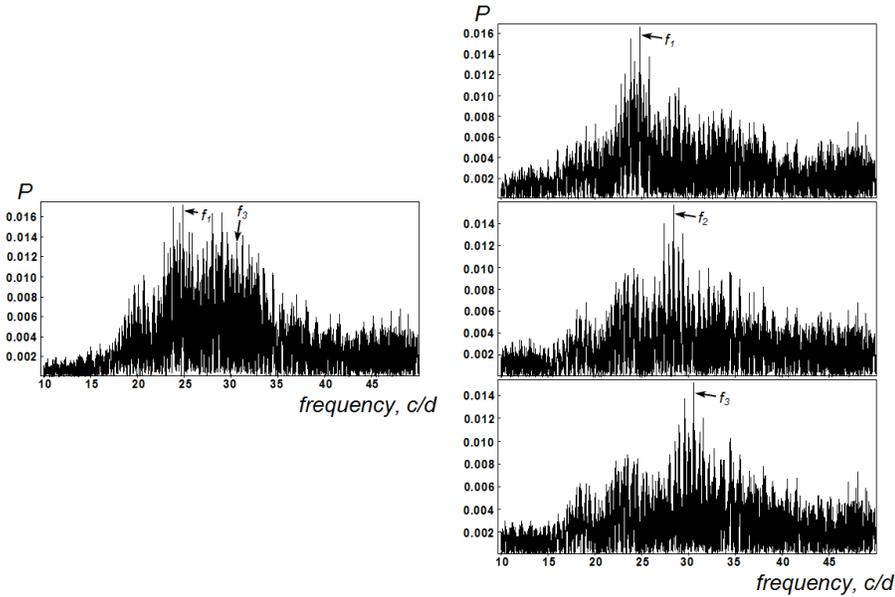}
\caption{Left:
the power spectrum of GSC 4560--02157. Right: The power spectra
for the first three frequencies, after all other frequencies were
pre-whitened}
\label{fig:5}
\end{figure}

The complex behavior of GSC 4560--02157 resembles that of
cataclysmic variables. To check the cataclysmic nature of the
star, we performed its spectroscopic observations.

The 6-m telescope of the Special Astrophysical Observatory (North
Caucasus, Russia) was used to take a single spectrum of the star
on 2014 June 5 UT 22:43:07 (JD hel 2456814.4466). The SCORPIO
camera with a VPHG 550g grism was used to take a spectrum of GSC
4560--02157 with a resolution of 13~\AA, the exposure time being
600~s. The computed heliocentric correction was $\Delta
V_r=-5.6$~km/s. The radial velocity of the star is nearly zero.
The spectrum (Fig.~6) reveals strong emission lines of the Balmer
series. HeI and HeII emissions are also clearly seen. The right
side of the spectrogram shows infrared interference fringes,
typical of CCD spectroscopy. Also present are features of
interstellar sodium and terrestrial water. The spectrogram
definitely confirms that GSC 4560--02157 belongs to cataclysmic
variables. Note that the spectrum was taken at the phase 0.505 of
the above light elements for the eclipsing variations, and thus at
approximately the secondary minimum. The $V$-band magnitudes of
the star measured just before the spectrogram, on JD 2456814.4395
and JD 2456814.4402, are 15.676 and 15.627. Thus, the star was
rather faint in the corresponding secondary minimum.

No X-ray source has so far been detected at the position of GSC
4560--2157, but this is no wonder for a faint and apparently
distant star.

\begin{figure*}
\centering
\includegraphics[width=15cm]{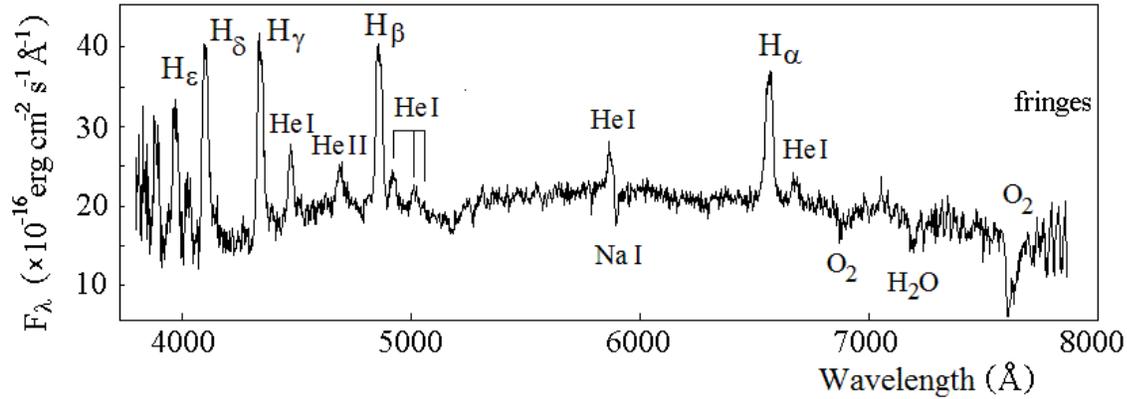}
\caption{The spectrum of GSC 4560--02157} \label{fig:6}
\end{figure*}

\section{Discussion}

The data we have accumulated for GSC 4560--02157 clearly identify
it as a cataclysmic eclipsing system. The Algol-type light curve
exhibits rapid start of the eclipse; the time from the beginning
of the eclipse to the onset of the totality phase is about
15~minutes. Thus, the eclipsed element of the system is rather
compact. The star's brightness is nearly constant during the
primary minimum, indicating that it is the eclipsed compact
element that is responsible for different brightness in different
orbital cycles and for multiperiodic short-term variations. The
latter could be pulsations of the white dwarf. The 0.04-day
pulsation period is rather large for pulsating white dwarfs
(ZZ~Ceti variables), but variables with such a behavior are known,
and some authors (e.g., Hermes et al. \cite{hermes}) suggest a new
type of extremely low-mass (ELM) white dwarfs. In our case,
several oscillations are clearly present, but their stability is
not beyond doubt; they can turn our to be quasi-periodic
oscillations, for example, like those noted for MT~Draconis
(Zubareva et al. \cite{zub}). The changing brightness level
outside eclipses is apparently related to the brightest, central
parts of the accretion disk, also occulted in primary eclipses.

\section{Conclusions}

The results of our photometric and spectroscopic observations
definitely show that GSC 4560--02157 is a new eclipsing
cataclysmic variable. So far, no outbursts have been detected for
the star, though it resembles known eclipsing dwarf novae both
photometrically and spectroscopically. Further observations are
needed to reveal possible additional manifestations of the star's
cataclysmic activity, like outbursts, high and low states.

\begin{acknowledgements}
We wish to thanks R.I.~Kokumbaeva, M.A. Krugov, N.V. Lichkanovsky,
I.V.~Rudakov, and W.~Mundrzyjewski for their assistance during the
observations. This study was partially supported by the Russian
Foundation for Basic Research (grants 12-02-31548, 13-02-00664,
13-02-00885, 14-02-00759), the programme of the President of
Russian Federation (grants MK-6686.2013.2, MK-1699.2014.2), the
Programme ``Non-stationary Phenomena of Objects in the Universe''
of the Presidium of Russian Academy of Sciences, and the program
``Studies of Physical Phenomena in Star-forming Regions and
Nuclear Zones of Active Galaxies'' (grant 0174/GF, Committee of
Science, Republic of Kazakhstan).
\end{acknowledgements}

\newpage

\begin{center}
{\bf Electronic supplement. Photometry of GSC 4560--02157 (Johnson $V$ band)}

\bigskip

\begin{tabular}{ccc|ccc|ccc}
   \hline
HJD & $V$, mag & Err &HJD & $V$, mag & Err &HJD & $V$, mag & Err\\
\hline
2456740.1422 & 15.726 & 0.017& 2456740.1967 & 15.439 & 0.021& 2456740.2529 & 15.586 & 0.013\\
2456740.1436 & 15.785 & 0.017& 2456740.1981 & 15.448 & 0.016& 2456740.2543 & 15.578 & 0.014\\
2456740.1450 & 15.860 & 0.017& 2456740.2010 & 15.424 & 0.016& 2456740.2556 & 15.616 & 0.015\\
2456740.1464 & 15.848 & 0.017& 2456740.2023 & 15.418 & 0.015& 2456740.2570 & 15.658 & 0.012\\
2456740.1478 & 15.791 & 0.020& 2456740.2037 & 15.432 & 0.017& 2456740.2599 & 15.766 & 0.015\\
2456740.1492 & 15.790 & 0.041& 2456740.2051 & 15.416 & 0.021& 2456740.2613 & 15.767 & 0.017\\
2456740.1506 & 15.746 & 0.044& 2456740.2065 & 15.369 & 0.017& 2456740.2627 & 15.856 & 0.018\\
\hline
   \end{tabular}
   \end{center}

\newpage
\begin{center}
{\bf Photometry of GSC 4560--02157 (Johnson $V$ band, continued)}

\bigskip

   \begin{tabular}{ccc|ccc|ccc}
   \hline
HJD & $V$, mag & Err &HJD & $V$, mag & Err &HJD & $V$, mag & Err\\
\hline
2456745.1430 & 15.368 & 0.007 & 2456745.1972 & 15.509 & 0.008 & 2456745.2529 & 15.257 & 0.003\\
2456745.1444 & 15.404 & 0.006 & 2456745.1986 & 15.502 & 0.008 & 2456745.2543 & 15.256 & 0.007\\
2456745.1457 & 15.430 & 0.006 & 2456745.2015 & 15.514 & 0.009 & 2456745.2557 & 15.303 & 0.007\\
2456745.1471 & 15.429 & 0.006 & 2456745.2029 & 15.514 & 0.008 & 2456745.2571 & 15.281 & 0.006\\
2456745.1485 & 15.395 & 0.007 & 2456745.2042 & 15.466 & 0.008 & 2456745.2585 & 15.211 & 0.006\\
2456745.1499 & 15.455 & 0.007 & 2456745.2056 & 15.445 & 0.008 & 2456745.2598 & 15.226 & 0.006\\
2456745.1513 & 15.486 & 0.006 & 2456745.2070 & 15.439 & 0.008 & 2456745.2612 & 15.247 & 0.003\\
\hline
   \end{tabular}
   \end{center}

\newpage
\begin{center}
{\bf Photometry of GSC 4560--02157 (Johnson $V$ band, continued)}

\bigskip

   \begin{tabular}{ccc|ccc|ccc}
   \hline
HJD & $V$, mag & Err &HJD & $V$, mag & Err &HJD & $V$, mag & Err\\
\hline
2456745.3071 & 15.870 & 0.007  & 2456745.3628 & 15.225 & 0.007 & 2456745.4171 & 15.335 & 0.006 \\
2456745.3085 & 15.914 & 0.007  & 2456745.3642 & 15.187 & 0.006 & 2456745.4185 & 15.390 & 0.006 \\
2456745.3099 & 15.953 & 0.007  & 2456745.3656 & 15.195 & 0.005 & 2456745.4199 & 15.401 & 0.017 \\
2456745.3128 & 15.977 & 0.009  & 2456745.3670 & 15.202 & 0.005 & 2456745.4212 & 15.356 & 0.019 \\
2456745.3142 & 15.959 & 0.007  & 2456745.3684 & 15.235 & 0.006 & 2456745.4355 & 15.463 & 0.018 \\
2456745.3155 & 15.965 & 0.009  & 2456745.3697 & 15.191 & 0.007 & 2456745.4368 & 15.492 & 0.017 \\
2456745.3169 & 15.972 & 0.010  & 2456745.3711 & 15.218 & 0.006 & 2456745.4382 & 15.520 & 0.014 \\
\hline
   \end{tabular}
   \end{center}

\newpage
\begin{center}
{\bf Photometry of GSC 4560--02157 (Johnson $V$ band, continued)}

\bigskip

   \begin{tabular}{ccc|ccc|ccc}
   \hline
HJD & $V$, mag & Err &HJD & $V$, mag & Err &HJD & $V$, mag & Err\\
\hline
2456745.4842 & 15.398 & 0.006  & 2456746.1402 & 15.549 & 0.009 & 2456746.1959 & 15.270 & 0.027\\
2456745.4856 & 15.404 & 0.006  & 2456746.1415 & 15.551 & 0.009 & 2456746.1987 & 15.282 & 0.027\\
2456745.4870 & 15.410 & 0.007  & 2456746.1429 & 15.533 & 0.009 & 2456746.2000 & 15.299 & 0.025\\
2456745.4884 & 15.407 & 0.009  & 2456746.1443 & 15.473 & 0.006 & 2456746.2014 & 15.337 & 0.012\\
2456745.4897 & 15.365 & 0.010  & 2456746.1457 & 15.448 & 0.009 & 2456746.2028 & 15.336 & 0.008\\
2456745.4911 & 15.336 & 0.009  & 2456746.1471 & 15.441 & 0.013 & 2456746.2042 & 15.337 & 0.007\\
2456745.4925 & 15.301 & 0.008  & 2456746.1485 & 15.447 & 0.018 & 2456746.2056 & 15.326 & 0.007\\
\hline
   \end{tabular}
   \end{center}

\newpage
\begin{center}
{\bf Photometry of GSC 4560--02157 (Johnson $V$ band, continued)}

\bigskip

   \begin{tabular}{ccc|ccc|ccc}
   \hline
HJD & $V$, mag & Err &HJD & $V$, mag & Err &HJD & $V$, mag & Err\\
\hline
2456771.2291 & 15.465 & 0.014  & 2456772.2247 & 15.443 & 0.007 & 2456772.3047 & 15.297 & 0.015\\
2456771.2313 & 15.421 & 0.014  & 2456772.2268 & 15.550 & 0.034 & 2456772.3069 & 15.367 & 0.014\\
2456771.2335 & 15.374 & 0.005  & 2456772.2289 & 15.559 & 0.038 & 2456772.3090 & 15.338 & 0.012\\
2456771.2356 & 15.390 & 0.004  & 2456772.2310 & 15.605 & 0.037 & 2456772.3111 & 15.194 & 0.013\\
2456771.2378 & 15.347 & 0.005  & 2456772.2332 & 15.563 & 0.036 & 2456772.3133 & 15.121 & 0.011\\
2456771.2400 & 15.295 & 0.007  & 2456772.2353 & 15.573 & 0.038 & 2456772.3154 & 15.227 & 0.013\\
2456771.2422 & 15.262 & 0.007  & 2456772.2374 & 15.575 & 0.029 & 2456772.3175 & 15.279 & 0.013\\
\hline
   \end{tabular}
   \end{center}

\newpage
\begin{center}
{\bf Photometry of GSC 4560--02157 (Johnson $V$ band, continued)}

\bigskip

   \begin{tabular}{ccc|ccc|ccc}
   \hline
HJD & $V$, mag & Err &HJD & $V$, mag & Err &HJD & $V$, mag & Err\\
\hline
2456772.3936 & 16.019 & 0.020  & 2456782.1868 & 15.408 & 0.013 & 2456782.2655 & 15.110 & 0.016\\
2456772.3957 & 15.989 & 0.018  & 2456782.1889 & 15.546 & 0.009 & 2456782.2677 & 15.060 & 0.012\\
2456772.3979 & 15.909 & 0.018  & 2456782.1910 & 15.675 & 0.006 & 2456782.2698 & 15.020 & 0.012\\
2456772.4000 & 15.747 & 0.016  & 2456782.1931 & 15.785 & 0.009 & 2456782.2719 & 15.034 & 0.007\\
2456772.4021 & 15.610 & 0.013  & 2456782.1953 & 15.880 & 0.010 & 2456782.2740 & 15.032 & 0.007\\
2456772.4043 & 15.477 & 0.015  & 2456782.1974 & 15.977 & 0.010 & 2456782.2762 & 15.007 & 0.008\\
2456772.4064 & 15.371 & 0.009  & 2456782.1995 & 16.021 & 0.011 & 2456782.2783 & 15.001 & 0.011\\
\hline
   \end{tabular}
   \end{center}

\newpage
\begin{center}
{\bf Photometry of GSC 4560--02157 (Johnson $V$ band, continued)}

\bigskip

   \begin{tabular}{ccc|ccc|ccc}
   \hline
HJD & $V$, mag & Err &HJD & $V$, mag & Err &HJD & $V$, mag & Err\\
\hline
2456782.3652 & 15.064 & 0.009  & 2456782.4440 & 15.208 & 0.017 & 2456784.2215 & 14.989 & 0.018\\
2456782.3674 & 15.049 & 0.009  & 2456782.4462 & 15.207 & 0.015 & 2456784.2235 & 14.998 & 0.018\\
2456782.3695 & 15.059 & 0.009  & 2456782.4483 & 15.169 & 0.016 & 2456784.2255 & 15.023 & 0.012\\
2456782.3716 & 15.058 & 0.009  & 2456782.4504 & 15.266 & 0.013 & 2456784.2275 & 15.012 & 0.008\\
2456782.3737 & 15.003 & 0.005  & 2456782.4525 & 15.388 & 0.013 & 2456784.2295 & 15.002 & 0.011\\
2456782.3759 & 14.986 & 0.006  & 2456784.1454 & 14.888 & 0.006 & 2456784.2315 & 15.011 & 0.015\\
2456782.3780 & 15.053 & 0.007  & 2456784.1476 & 14.922 & 0.006 & 2456784.2335 & 14.944 & 0.016\\
\hline
   \end{tabular}
   \end{center}

\newpage
\begin{center}
{\bf Photometry of GSC 4560--02157 (Johnson $V$ band, continued)}

\bigskip

   \begin{tabular}{ccc|ccc|ccc}
   \hline
HJD & $V$, mag & Err &HJD & $V$, mag & Err &HJD & $V$, mag & Err\\
\hline
2456784.2990 & 14.947 & 0.005  & 2456784.3729 & 15.076 & 0.017 & 2456784.4091 & 14.812 & 0.007\\
2456784.3011 & 14.995 & 0.006  & 2456784.3750 & 15.029 & 0.013 & 2456784.4113 & 14.765 & 0.017\\
2456784.3033 & 14.992 & 0.009  & 2456784.3771 & 15.070 & 0.012 & 2456784.4134 & 14.821 & 0.017\\
2456784.3054 & 15.027 & 0.011  & 2456784.3792 & 15.066 & 0.011 & 2456784.4156 & 14.748 & 0.014\\
2456784.3075 & 15.113 & 0.010  & 2456784.3813 & 15.021 & 0.013 & 2456784.4177 & 14.841 & 0.013\\
2456784.3097 & 15.217 & 0.010  & 2456784.3835 & 15.028 & 0.013 & 2456784.4199 & 14.887 & 0.013\\
2456784.3118 & 15.316 & 0.010  & 2456784.3857 & 14.996 & 0.011 & 2456784.4220 & 14.852 & 0.008\\
2456784.3139 & 15.520 & 0.008  & 2456784.3878 & 14.963 & 0.011 & 2456784.4241 & 14.900 & 0.009\\
2456784.3161 & 15.619 & 0.009  & 2456784.3899 & 14.874 & 0.011 & 2456784.4263 & 14.859 & 0.009\\
2456784.3182 & 15.769 & 0.010  & 2456784.3920 & 14.880 & 0.012 & 2456784.4284 & 14.877 & 0.020\\
2456784.3528 & 15.127 & 0.011  & 2456784.3942 & 14.853 & 0.012 & 2456784.4369 & 14.971 & 0.019\\
2456784.3564 & 15.066 & 0.013  & 2456784.3963 & 14.811 & 0.011 & 2456784.4390 & 14.977 & 0.015\\
2456784.3586 & 15.109 & 0.013  & 2456784.3985 & 14.833 & 0.009 & 2456784.4452 & 14.928 & 0.018\\
2456784.3607 & 15.101 & 0.019  & 2456784.4006 & 14.821 & 0.008 & 2456784.4472 & 14.909 & 0.016\\
2456784.3628 & 15.053 & 0.020  & 2456784.4028 & 14.871 & 0.004 & 2456784.4492 & 14.894 & 0.015\\
2456784.3669 & 15.096 & 0.021  & 2456784.4049 & 14.864 & 0.007 & 2456784.4512 & 14.950 & 0.014\\
2456784.3709 & 15.028 & 0.017  & 2456784.4070 & 14.844 & 0.006 & 2456784.4532 & 14.928 & 0.014\\
\hline
   \end{tabular}
   \end{center}

\newpage

\begin{center}
{\bf Electronic supplement. Photometry of GSC 4560--02157 (Johnson $R$ band)}

\bigskip

   \begin{tabular}{ccc|ccc|ccc}
   \hline
HJD & $R$, mag & Err &HJD & $R$, mag & Err &HJD & $R$, mag & Err\\
\hline
2456364.4342 & 14.974 & 0.005 & 2456371.4766 & 15.065 & 0.011 & 2456374.3476 & 15.463 & 0.006\\
2456364.4353 & 14.971 & 0.005 & 2456371.4777 & 15.054 & 0.011 & 2456374.3488 & 15.453 & 0.011\\
2456364.4365 & 15.006 & 0.005 & 2456371.4789 & 15.043 & 0.009 & 2456374.3500 & 15.455 & 0.010\\
2456364.4376 & 14.996 & 0.006 & 2456371.4801 & 15.036 & 0.009 & 2456374.3512 & 15.433 & 0.012\\
2456364.4387 & 15.002 & 0.011 & 2456371.4813 & 15.024 & 0.006 & 2456374.3524 & 15.450 & 0.010\\
2456364.4398 & 15.000 & 0.016 & 2456371.4987 & 15.030 & 0.006 & 2456374.3535 & 15.409 & 0.011\\
2456364.4410 & 15.019 & 0.026 & 2456371.5024 & 15.030 & 0.009 & 2456374.3547 & 15.419 & 0.009\\
\hline
   \end{tabular}
   \end{center}

\newpage

\begin{center}
{\bf Photometry of GSC 4560--02157 (Johnson $R$ band, continued)}

\bigskip

   \begin{tabular}{ccc|ccc|ccc}
   \hline
HJD & $R$, mag & Err &HJD & $R$, mag & Err &HJD & $R$, mag & Err\\
\hline
2456585.0804 & 14.936 & 0.013 & 2456588.0642 & 15.068 & 0.004 & 2456591.0305 & 15.201 & 0.008\\
2456585.0818 & 14.879 & 0.009 & 2456588.0656 & 15.054 & 0.004 & 2456591.0319 & 15.171 & 0.008\\
2456585.0841 & 14.871 & 0.010 & 2456588.0670 & 15.073 & 0.005 & 2456591.0333 & 15.164 & 0.006\\
2456585.0858 & 14.877 & 0.007 & 2456588.0684 & 15.030 & 0.004 & 2456591.0347 & 15.135 & 0.003\\
2456585.0875 & 14.910 & 0.007 & 2456588.0698 & 15.061 & 0.004 & 2456591.0361 & 15.155 & 0.003\\
2456585.0893 & 14.956 & 0.008 & 2456588.0711 & 15.079 & 0.004 & 2456591.0375 & 15.131 & 0.004\\
2456585.0910 & 14.972 & 0.009 & 2456588.0725 & 15.111 & 0.005 & 2456591.0388 & 15.134 & 0.005\\
\hline
   \end{tabular}
   \end{center}

\newpage

\begin{center}
{\bf Photometry of GSC 4560--02157 (Johnson $R$ band, continued)}

\bigskip

   \begin{tabular}{ccc|ccc|ccc}
   \hline
HJD & $R$, mag & Err &HJD & $R$, mag & Err &HJD & $R$, mag & Err\\
\hline
2456591.0817 & 14.849 & 0.005 & 2456591.1330 & 15.361 & 0.009 & 2456592.0746 & 15.105 & 0.005\\
2456591.0831 & 14.834 & 0.008 & 2456591.1344 & 15.394 & 0.009 & 2456592.0760 & 15.097 & 0.005\\
2456591.0845 & 14.808 & 0.008 & 2456591.1358 & 15.399 & 0.008 & 2456592.0774 & 15.103 & 0.006\\
2456591.0859 & 14.816 & 0.008 & 2456591.1371 & 15.411 & 0.008 & 2456592.0788 & 15.099 & 0.006\\
2456591.0873 & 14.833 & 0.007 & 2456591.1385 & 15.412 & 0.007 & 2456592.0801 & 15.109 & 0.005\\
2456591.0887 & 14.817 & 0.008 & 2456591.1399 & 15.410 & 0.007 & 2456592.0815 & 15.132 & 0.004\\
2456591.0900 & 14.817 & 0.008 & 2456591.1413 & 15.433 & 0.008 & 2456592.0829 & 15.113 & 0.003\\
\hline
   \end{tabular}
   \end{center}

\newpage

\begin{center}
{\bf Photometry of GSC 4560--02157 (Johnson $R$ band, continued)}

\bigskip

   \begin{tabular}{ccc|ccc|ccc}
   \hline
HJD & $R$, mag & Err &HJD & $R$, mag & Err &HJD & $R$, mag & Err\\
\hline
2456594.0462 & 14.994 & 0.013 & 2456594.0973 & 14.925 & 0.006 & 2456594.1484 & 14.801 & 0.011\\
2456594.0476 & 15.044 & 0.013 & 2456594.0987 & 14.926 & 0.006 & 2456594.1498 & 14.798 & 0.009\\
2456594.0489 & 15.114 & 0.012 & 2456594.1001 & 14.899 & 0.006 & 2456594.1512 & 14.812 & 0.008\\
2456594.0503 & 15.201 & 0.009 & 2456594.1015 & 14.878 & 0.008 & 2456594.1526 & 14.825 & 0.007\\
2456594.0517 & 15.254 & 0.007 & 2456594.1028 & 14.889 & 0.006 & 2456594.1540 & 14.818 & 0.007\\
2456594.0531 & 15.307 & 0.007 & 2456594.1042 & 14.884 & 0.006 & 2456594.1554 & 14.837 & 0.009\\
2456594.0545 & 15.373 & 0.010 & 2456594.1056 & 14.894 & 0.007 & 2456594.1567 & 14.839 & 0.010\\
\hline
   \end{tabular}
   \end{center}

\newpage

\begin{center}
{\bf Photometry of GSC 4560--02157 (Johnson $R$ band, continued)}

\bigskip

   \begin{tabular}{ccc|ccc|ccc}
   \hline
HJD & $R$, mag & Err &HJD & $R$, mag & Err &HJD & $R$, mag & Err\\
\hline
2456598.0760 & 14.906 & 0.031 & 2456716.3851 & 15.391 & 0.006 & 2456716.4341 & 14.958 & 0.003\\
2456598.0773 & 14.880 & 0.035 & 2456716.3864 & 15.416 & 0.004 & 2456716.4354 & 14.973 & 0.002\\
2456598.0787 & 14.835 & 0.030 & 2456716.3877 & 15.437 & 0.005 & 2456716.4368 & 14.973 & 0.002\\
2456598.0801 & 14.847 & 0.030 & 2456716.3890 & 15.437 & 0.005 & 2456716.4381 & 14.980 & 0.005\\
2456716.3413 & 14.823 & 0.005 & 2456716.3904 & 15.446 & 0.002 & 2456716.4394 & 14.947 & 0.006\\
2456716.3426 & 14.829 & 0.005 & 2456716.3917 & 15.447 & 0.002 & 2456716.4407 & 14.933 & 0.007\\
2456716.3440 & 14.827 & 0.005 & 2456716.3930 & 15.439 & 0.003 & 2456716.4421 & 14.930 & 0.006\\
\hline
   \end{tabular}
   \end{center}

\newpage

\begin{center}
{\bf Photometry of GSC 4560--02157 (Johnson $R$ band, continued)}

\bigskip

   \begin{tabular}{ccc|ccc|ccc}
   \hline
HJD & $R$, mag & Err &HJD & $R$, mag & Err &HJD & $R$, mag & Err\\
\hline
2456716.4832 & 14.845 & 0.003 & 2456717.1981 & 15.419 & 0.006 & 2456717.2471 & 14.814 & 0.006\\
2456716.4845 & 14.883 & 0.007 & 2456717.1994 & 15.388 & 0.005 & 2456717.2484 & 14.824 & 0.006\\
2456716.4858 & 14.902 & 0.007 & 2456717.2007 & 15.360 & 0.004 & 2456717.2498 & 14.816 & 0.005\\
2456716.4871 & 14.924 & 0.007 & 2456717.2020 & 15.332 & 0.005 & 2456717.2511 & 14.818 & 0.005\\
2456716.4885 & 14.942 & 0.008 & 2456717.2034 & 15.266 & 0.005 & 2456717.2524 & 14.788 & 0.004\\
2456716.4898 & 14.941 & 0.007 & 2456717.2047 & 15.224 & 0.004 & 2456717.2537 & 14.803 & 0.004\\
2456716.4911 & 14.956 & 0.005 & 2456717.2060 & 15.151 & 0.005 & 2456717.2551 & 14.800 & 0.005\\
\hline
   \end{tabular}
   \end{center}

\newpage

\begin{center}
{\bf Photometry of GSC 4560--02157 (Johnson $R$ band, continued)}

\bigskip

   \begin{tabular}{ccc|ccc|ccc}
   \hline
HJD & $R$, mag & Err &HJD & $R$, mag & Err &HJD & $R$, mag & Err\\
\hline
2456717.2961 & 14.782 & 0.007 & 2456717.3500 & 14.712 & 0.009 & 2456717.4014 & 14.641 & 0.005\\
2456717.2974 & 14.791 & 0.007 & 2456717.3514 & 14.721 & 0.008 & 2456717.4028 & 14.682 & 0.006\\
2456717.2988 & 14.834 & 0.006 & 2456717.3528 & 14.718 & 0.008 & 2456717.4042 & 14.710 & 0.006\\
2456717.3001 & 14.865 & 0.007 & 2456717.3542 & 14.699 & 0.007 & 2456717.4056 & 14.711 & 0.006\\
2456717.3014 & 14.863 & 0.003 & 2456717.3555 & 14.706 & 0.005 & 2456717.4070 & 14.733 & 0.006\\
2456717.3028 & 14.897 & 0.003 & 2456717.3569 & 14.727 & 0.004 & 2456717.4084 & 14.777 & 0.003\\
2456717.3041 & 14.887 & 0.005 & 2456717.3583 & 14.719 & 0.002 & 2456717.4097 & 14.770 & 0.003\\
\hline
   \end{tabular}
   \end{center}

\newpage

\begin{center}
{\bf Photometry of GSC 4560--02157 (Johnson $R$ band, continued)}

\bigskip

   \begin{tabular}{ccc|ccc|ccc}
   \hline
HJD & $R$, mag & Err &HJD & $R$, mag & Err &HJD & $R$, mag & Err\\
\hline
2456717.4528 & 15.434 & 0.004 & 2456717.5110 & 14.737 & 0.006 & 2456718.2584 & 15.434 & 0.002\\
2456717.4542 & 15.442 & 0.004 & 2456717.5124 & 14.752 & 0.006 & 2456718.2597 & 15.414 & 0.002\\
2456717.4555 & 15.428 & 0.005 & 2456717.5137 & 14.755 & 0.004 & 2456718.2611 & 15.401 & 0.002\\
2456717.4569 & 15.433 & 0.005 & 2456717.5150 & 14.744 & 0.005 & 2456718.2625 & 15.361 & 0.003\\
2456717.4583 & 15.441 & 0.003 & 2456717.5163 & 14.730 & 0.005 & 2456718.2639 & 15.318 & 0.004\\
2456717.4597 & 15.440 & 0.004 & 2456717.5177 & 14.711 & 0.004 & 2456718.2653 & 15.251 & 0.004\\
2456717.4611 & 15.438 & 0.004 & 2456717.5190 & 14.730 & 0.004 & 2456718.2667 & 15.192 & 0.004\\
\hline
   \end{tabular}
   \end{center}

\newpage

\begin{center}
{\bf Photometry of GSC 4560--02157 (Johnson $R$ band, continued)}

\bigskip

   \begin{tabular}{ccc|ccc|ccc}
   \hline
HJD & $R$, mag & Err &HJD & $R$, mag & Err &HJD & $R$, mag & Err\\
\hline
2456718.3096 & 14.753 & 0.006 & 2456718.3608 & 14.846 & 0.006 & 2456718.4121 & 14.788 & 0.009\\
2456718.3110 & 14.772 & 0.005 & 2456718.3622 & 14.846 & 0.008 & 2456718.4134 & 14.779 & 0.005\\
2456718.3124 & 14.763 & 0.006 & 2456718.3636 & 14.847 & 0.008 & 2456718.4148 & 14.761 & 0.005\\
2456718.3138 & 14.777 & 0.005 & 2456718.3650 & 14.878 & 0.003 & 2456718.4162 & 14.734 & 0.003\\
2456718.3152 & 14.786 & 0.006 & 2456718.3664 & 14.898 & 0.003 & 2456718.4176 & 14.713 & 0.004\\
2456718.3165 & 14.762 & 0.006 & 2456718.3678 & 14.899 & 0.003 & 2456718.4190 & 14.741 & 0.003\\
2456718.3179 & 14.773 & 0.006 & 2456718.3692 & 14.896 & 0.007 & 2456718.4204 & 14.741 & 0.004\\
\hline
   \end{tabular}
   \end{center}

\newpage

\begin{center}
{\bf Photometry of GSC 4560--02157 (Johnson $R$ band, continued)}

\bigskip

   \begin{tabular}{ccc|ccc|ccc}
   \hline
HJD & $R$, mag & Err &HJD & $R$, mag & Err &HJD & $R$, mag & Err\\
\hline
2456718.4633 & 14.696 & 0.003 & 2456718.5325 & 15.123 & 0.005 & 2456719.2059 & 14.726 & 0.008\\
2456718.4647 & 14.692 & 0.004 & 2456719.1546 & 14.810 & 0.003 & 2456719.2073 & 14.755 & 0.008\\
2456718.4660 & 14.700 & 0.005 & 2456719.1560 & 14.816 & 0.003 & 2456719.2087 & 14.802 & 0.008\\
2456718.4674 & 14.705 & 0.005 & 2456719.1574 & 14.797 & 0.003 & 2456719.2101 & 14.805 & 0.007\\
2456718.4688 & 14.715 & 0.004 & 2456719.1588 & 14.802 & 0.002 & 2456719.2115 & 14.808 & 0.008\\
2456718.4702 & 14.723 & 0.005 & 2456719.1602 & 14.836 & 0.002 & 2456719.2129 & 14.783 & 0.005\\
2456718.4716 & 14.723 & 0.004 & 2456719.1616 & 14.841 & 0.002 & 2456719.2143 & 14.737 & 0.005\\
\hline
   \end{tabular}
   \end{center}

\newpage

\begin{center}
{\bf Photometry of GSC 4560--02157 (Johnson $R$ band, continued)}

\bigskip

   \begin{tabular}{ccc|ccc|ccc}
   \hline
HJD & $R$, mag & Err &HJD & $R$, mag & Err &HJD & $R$, mag & Err\\
\hline
2456719.2572 & 14.612 & 0.004 & 2456719.3099 & 15.436 & 0.005 & 2456719.3612 & 14.822 & 0.005\\
2456719.2586 & 14.622 & 0.006 & 2456719.3113 & 15.441 & 0.005 & 2456719.3626 & 14.822 & 0.004\\
2456719.2600 & 14.635 & 0.006 & 2456719.3127 & 15.447 & 0.005 & 2456719.3639 & 14.834 & 0.003\\
2456719.2614 & 14.621 & 0.006 & 2456719.3141 & 15.441 & 0.005 & 2456719.3653 & 14.837 & 0.003\\
2456719.2641 & 14.616 & 0.006 & 2456719.3154 & 15.443 & 0.003 & 2456719.3667 & 14.830 & 0.003\\
2456719.2655 & 14.628 & 0.005 & 2456719.3168 & 15.440 & 0.006 & 2456719.3681 & 14.795 & 0.002\\
2456719.2669 & 14.620 & 0.002 & 2456719.3182 & 15.424 & 0.006 & 2456719.3695 & 14.774 & 0.002\\
\hline
   \end{tabular}
   \end{center}

\newpage

\begin{center}
{\bf Photometry of GSC 4560--02157 (Johnson $R$ band, continued)}

\bigskip

   \begin{tabular}{ccc|ccc|ccc}
   \hline
HJD & $R$, mag & Err &HJD & $R$, mag & Err &HJD & $R$, mag & Err\\
\hline
2456719.4124 & 14.857 & 0.002 & 2456719.4636 & 14.924 & 0.005 & 2456720.1876 & 14.803 & 0.008\\
2456719.4138 & 14.852 & 0.003 & 2456719.4650 & 14.907 & 0.005 & 2456720.1890 & 14.798 & 0.008\\
2456719.4151 & 14.821 & 0.003 & 2456719.4664 & 14.928 & 0.005 & 2456720.1903 & 14.758 & 0.008\\
2456719.4165 & 14.801 & 0.004 & 2456719.4677 & 14.886 & 0.003 & 2456720.1917 & 14.775 & 0.003\\
2456719.4179 & 14.850 & 0.005 & 2456719.4691 & 14.869 & 0.004 & 2456720.1931 & 14.787 & 0.003\\
2456719.4193 & 14.868 & 0.006 & 2456719.4705 & 14.891 & 0.005 & 2456720.1945 & 14.780 & 0.005\\
2456719.4207 & 14.889 & 0.005 & 2456719.4719 & 14.885 & 0.005 & 2456720.1959 & 14.756 & 0.004\\
\hline
   \end{tabular}
   \end{center}

\newpage

\begin{center}
{\bf Photometry of GSC 4560--02157 (Johnson $R$ band, continued)}

\bigskip

   \begin{tabular}{ccc|ccc|ccc}
   \hline
HJD & $R$, mag & Err &HJD & $R$, mag & Err &HJD & $R$, mag & Err\\
\hline
2456720.2387 & 14.932 & 0.004 & 2456720.2899 & 14.732 & 0.003 & 2456720.3411 & 14.790 & 0.006\\
2456720.2401 & 14.937 & 0.002 & 2456720.2913 & 14.715 & 0.003 & 2456720.3425 & 14.777 & 0.006\\
2456720.2415 & 14.942 & 0.002 & 2456720.2926 & 14.732 & 0.004 & 2456720.3439 & 14.803 & 0.006\\
2456720.2429 & 14.962 & 0.002 & 2456720.2940 & 14.710 & 0.004 & 2456720.3452 & 14.879 & 0.006\\
2456720.2442 & 14.960 & 0.002 & 2456720.2954 & 14.747 & 0.005 & 2456720.3466 & 14.894 & 0.002\\
2456720.2456 & 14.977 & 0.003 & 2456720.2968 & 14.777 & 0.004 & 2456720.3480 & 14.915 & 0.002\\
2456720.2470 & 14.988 & 0.003 & 2456720.2982 & 14.758 & 0.002 & 2456720.3494 & 14.954 & 0.002\\
\hline
   \end{tabular}
   \end{center}

\newpage

\begin{center}
{\bf Photometry of GSC 4560--02157 (Johnson $R$ band, continued)}

\bigskip

   \begin{tabular}{ccc|ccc|ccc}
   \hline
HJD & $R$, mag & Err &HJD & $R$, mag & Err &HJD & $R$, mag & Err\\
\hline
2456720.3923 & 15.012 & 0.004 & 2456720.4444 & 14.687 & 0.002 & 2456720.5072 & 15.016 & 0.007\\
2456720.3937 & 14.888 & 0.003 & 2456720.4458 & 14.688 & 0.002 & 2456720.5086 & 15.060 & 0.008\\
2456720.3950 & 14.862 & 0.006 & 2456720.4472 & 14.690 & 0.002 & 2456720.5099 & 15.062 & 0.010\\
2456720.3964 & 14.887 & 0.006 & 2456720.4486 & 14.713 & 0.002 & 2456720.5112 & 15.050 & 0.008\\
2456720.3978 & 14.912 & 0.006 & 2456720.4500 & 14.673 & 0.002 & 2456720.5125 & 15.057 & 0.008\\
2456720.3992 & 14.909 & 0.006 & 2456720.4514 & 14.698 & 0.003 & 2456720.5139 & 14.967 & 0.008\\
2456720.4006 & 14.895 & 0.005 & 2456720.4527 & 14.739 & 0.003 & 2456720.5152 & 15.008 & 0.008\\
\hline
   \end{tabular}
   \end{center}

\newpage

\begin{center}
{\bf Photometry of GSC 4560--02157 (Johnson $R$ band, continued)}

\bigskip

   \begin{tabular}{ccc|ccc|ccc}
   \hline
HJD & $R$, mag & Err &HJD & $R$, mag & Err &HJD & $R$, mag & Err\\
\hline
2456722.3016 & 14.904 & 0.002 & 2456722.3528 & 15.219 & 0.004 & 2456722.4041 & 14.761 & 0.007\\
2456722.3029 & 14.891 & 0.002 & 2456722.3542 & 15.205 & 0.004 & 2456722.4055 & 14.751 & 0.006\\
2456722.3043 & 14.914 & 0.002 & 2456722.3556 & 15.216 & 0.004 & 2456722.4069 & 14.756 & 0.007\\
2456722.3057 & 14.912 & 0.003 & 2456722.3570 & 15.215 & 0.005 & 2456722.4083 & 14.762 & 0.006\\
2456722.3071 & 14.937 & 0.004 & 2456722.3584 & 15.203 & 0.006 & 2456722.4097 & 14.814 & 0.004\\
2456722.3085 & 14.961 & 0.003 & 2456722.3598 & 15.198 & 0.006 & 2456722.4111 & 14.825 & 0.005\\
2456722.3099 & 14.952 & 0.003 & 2456722.3612 & 15.145 & 0.006 & 2456722.4125 & 14.830 & 0.003\\
\hline
   \end{tabular}
   \end{center}

\newpage

\begin{center}
{\bf Photometry of GSC 4560--02157 (Johnson $R$ band, continued)}

\bigskip

   \begin{tabular}{ccc|ccc|ccc}
   \hline
HJD & $R$, mag & Err &HJD & $R$, mag & Err &HJD & $R$, mag & Err\\
\hline
2456722.4554 & 14.902 & 0.003 & 2456722.5129 & 15.288 & 0.005 & 2456728.1680 & 15.122 & 0.009\\
2456722.4568 & 14.946 & 0.002 & 2456722.5142 & 15.232 & 0.005 & 2456728.1692 & 15.120 & 0.005\\
2456722.4582 & 14.978 & 0.002 & 2456722.5156 & 15.159 & 0.006 & 2456728.1704 & 15.111 & 0.009\\
2456722.4596 & 14.976 & 0.002 & 2456722.5169 & 15.119 & 0.003 & 2456728.1717 & 15.076 & 0.009\\
2456722.4610 & 14.981 & 0.003 & 2456722.5182 & 15.121 & 0.003 & 2456728.1729 & 15.078 & 0.007\\
2456722.4624 & 14.991 & 0.003 & 2456722.5195 & 15.145 & 0.004 & 2456728.1741 & 15.111 & 0.006\\
2456722.4637 & 14.975 & 0.003 & 2456722.5209 & 15.145 & 0.004 & 2456728.1753 & 15.136 & 0.006\\
\hline
   \end{tabular}
   \end{center}

\newpage

\begin{center}
{\bf Photometry of GSC 4560--02157 (Johnson $R$ band, continued)}

\bigskip

   \begin{tabular}{ccc|ccc|ccc}
   \hline
HJD & $R$, mag & Err &HJD & $R$, mag & Err &HJD & $R$, mag & Err\\
\hline
2456728.2134 & 15.133 & 0.007 & 2456728.2588 & 14.878 & 0.003 & 2456728.3042 & 15.060 & 0.007\\
2456728.2146 & 15.083 & 0.002 & 2456728.2600 & 14.891 & 0.002 & 2456728.3054 & 15.075 & 0.007\\
2456728.2159 & 15.131 & 0.004 & 2456728.2613 & 14.885 & 0.002 & 2456728.3067 & 15.080 & 0.006\\
2456728.2171 & 15.135 & 0.004 & 2456728.2625 & 14.859 & 0.002 & 2456728.3079 & 15.088 & 0.003\\
2456728.2183 & 15.142 & 0.004 & 2456728.2637 & 14.882 & 0.002 & 2456728.3091 & 15.101 & 0.003\\
2456728.2195 & 15.133 & 0.002 & 2456728.2649 & 14.895 & 0.002 & 2456728.3104 & 15.104 & 0.002\\
2456728.2208 & 15.121 & 0.002 & 2456728.2662 & 14.910 & 0.002 & 2456728.3116 & 15.121 & 0.003\\
\hline
   \end{tabular}
   \end{center}

\newpage

\begin{center}
{\bf Photometry of GSC 4560--02157 (Johnson $R$ band, continued)}

\bigskip

   \begin{tabular}{ccc|ccc|ccc}
   \hline
HJD & $R$, mag & Err &HJD & $R$, mag & Err &HJD & $R$, mag & Err\\
\hline
2456728.3496 & 15.332 & 0.005 & 2456728.3951 & 14.980 & 0.005 & 2456728.4405 & 15.068 & 0.005\\
2456728.3509 & 15.287 & 0.004 & 2456728.3963 & 14.960 & 0.004 & 2456728.4417 & 15.098 & 0.004\\
2456728.3521 & 15.247 & 0.005 & 2456728.3975 & 14.970 & 0.003 & 2456728.4430 & 15.114 & 0.004\\
2456728.3533 & 15.216 & 0.006 & 2456728.3988 & 14.975 & 0.002 & 2456728.4442 & 15.131 & 0.003\\
2456728.3546 & 15.202 & 0.005 & 2456728.4000 & 14.970 & 0.007 & 2456728.4454 & 15.143 & 0.003\\
2456728.3558 & 15.175 & 0.004 & 2456728.4012 & 15.000 & 0.007 & 2456728.4466 & 15.140 & 0.006\\
2456728.3570 & 15.172 & 0.005 & 2456728.4024 & 15.004 & 0.006 & 2456728.4479 & 15.164 & 0.005\\
\hline
   \end{tabular}
   \end{center}

\newpage

\begin{center}
{\bf Photometry of GSC 4560--02157 (Johnson $R$ band, continued)}

\bigskip

   \begin{tabular}{ccc|ccc|ccc}
   \hline
HJD & $R$, mag & Err &HJD & $R$, mag & Err &HJD & $R$, mag & Err\\
\hline
2456728.4989 & 15.096 & 0.003 & 2456745.3335 & 15.172 & 0.005 & 2456771.2258 & 14.978 & 0.009\\
2456728.5001 & 15.089 & 0.004 & 2456745.3558 & 14.888 & 0.004 & 2456771.2279 & 14.973 & 0.004\\
2456728.5013 & 15.065 & 0.004 & 2456745.3781 & 14.887 & 0.007 & 2456771.2301 & 14.965 & 0.004\\
2456728.5024 & 15.058 & 0.004 & 2456745.4004 & 14.858 & 0.008 & 2456771.2323 & 14.939 & 0.004\\
2456728.5036 & 15.035 & 0.004 & 2456745.4340 & 14.954 & 0.008 & 2456771.2345 & 14.920 & 0.005\\
2456728.5048 & 15.036 & 0.004 & 2456745.4563 & 15.072 & 0.008 & 2456771.2367 & 14.899 & 0.009\\
2456728.5059 & 15.026 & 0.002 & 2456745.4786 & 14.977 & 0.008 & 2456771.2389 & 14.863 & 0.011\\
\hline
   \end{tabular}
   \end{center}

\newpage

\begin{center}
{\bf Photometry of GSC 4560--02157 (Johnson $R$ band, continued)}

\bigskip

   \begin{tabular}{ccc|ccc|ccc}
   \hline
HJD & $R$, mag & Err &HJD & $R$, mag & Err &HJD & $R$, mag & Err\\
\hline
2456771.2258 & 14.978 & 0.009 & 2456772.2235 & 14.999 & 0.012 & 2456772.3036 & 14.929 & 0.005\\
2456771.2279 & 14.973 & 0.004 & 2456772.2256 & 15.076 & 0.011 & 2456772.3057 & 14.889 & 0.009\\
2456771.2301 & 14.965 & 0.004 & 2456772.2278 & 15.117 & 0.011 & 2456772.3078 & 14.945 & 0.011\\
2456771.2323 & 14.939 & 0.004 & 2456772.2299 & 15.169 & 0.011 & 2456772.3100 & 14.859 & 0.016\\
2456771.2345 & 14.920 & 0.005 & 2456772.2320 & 15.169 & 0.005 & 2456772.3121 & 14.788 & 0.016\\
2456771.2367 & 14.899 & 0.009 & 2456772.2341 & 15.162 & 0.003 & 2456772.3142 & 14.809 & 0.017\\
2456771.2389 & 14.863 & 0.011 & 2456772.2363 & 15.188 & 0.003 & 2456772.3164 & 14.822 & 0.015\\
\hline
   \end{tabular}
   \end{center}

\newpage

\begin{center}
{\bf Photometry of GSC 4560--02157 (Johnson $R$ band, continued)}

\bigskip

   \begin{tabular}{ccc|ccc|ccc}
   \hline
HJD & $R$, mag & Err &HJD & $R$, mag & Err &HJD & $R$, mag & Err\\
\hline
2456772.3946 & 15.490 & 0.005 & 2456782.1856 & 14.853 & 0.011 & 2456782.2644 & 14.682 & 0.009\\
2456772.3967 & 15.452 & 0.005 & 2456782.1877 & 14.992 & 0.013 & 2456782.2665 & 14.654 & 0.013\\
2456772.3989 & 15.362 & 0.004 & 2456782.1899 & 15.132 & 0.013 & 2456782.2686 & 14.631 & 0.014\\
2456772.4010 & 15.258 & 0.004 & 2456782.1920 & 15.226 & 0.013 & 2456782.2708 & 14.642 & 0.013\\
2456772.4031 & 15.104 & 0.005 & 2456782.1941 & 15.298 & 0.010 & 2456782.2729 & 14.624 & 0.012\\
2456772.4052 & 15.027 & 0.007 & 2456782.1962 & 15.375 & 0.009 & 2456782.2750 & 14.648 & 0.011\\
2456772.4074 & 14.965 & 0.007 & 2456782.1984 & 15.409 & 0.007 & 2456782.2772 & 14.568 & 0.009\\
\hline
   \end{tabular}
   \end{center}

\newpage

\begin{center}
{\bf Photometry of GSC 4560--02157 (Johnson $R$ band, continued)}

\bigskip

   \begin{tabular}{ccc|ccc|ccc}
   \hline
HJD & $R$, mag & Err &HJD & $R$, mag & Err &HJD & $R$, mag & Err\\
\hline
2456782.3662 & 14.664 & 0.010 & 2456782.4450 & 14.756 & 0.018 & 2456784.2184 & 14.585 & 0.059\\
2456782.3683 & 14.661 & 0.010 & 2456782.4471 & 14.789 & 0.017 & 2456784.2204 & 14.563 & 0.006\\
2456782.3705 & 14.677 & 0.015 & 2456782.4493 & 14.814 & 0.012 & 2456784.2224 & 14.574 & 0.007\\
2456782.3726 & 14.621 & 0.009 & 2456782.4514 & 14.880 & 0.013 & 2456784.2244 & 14.582 & 0.008\\
2456782.3747 & 14.576 & 0.008 & 2456782.4535 & 14.955 & 0.013 & 2456784.2264 & 14.587 & 0.008\\
2456782.3768 & 14.595 & 0.008 & 2456784.1464 & 14.487 & 0.021 & 2456784.2284 & 14.596 & 0.006\\
2456782.3790 & 14.656 & 0.008 & 2456784.1486 & 14.525 & 0.021 & 2456784.2304 & 14.599 & 0.006\\
\hline
   \end{tabular}
   \end{center}

\newpage

\begin{center}
{\bf Photometry of GSC 4560--02157 (Johnson $R$ band, continued)}

\bigskip

   \begin{tabular}{ccc|ccc|ccc}
   \hline
HJD & $R$, mag & Err &HJD & $R$, mag & Err &HJD & $R$, mag & Err\\
\hline
2456784.2957 & 14.492 & 0.011 & 2456784.3739 & 14.668 & 0.025 & 2456784.4101 & 14.400 & 0.015\\
2456784.2979 & 14.540 & 0.006 & 2456784.3759 & 14.641 & 0.007 & 2456784.4123 & 14.415 & 0.015\\
2456784.3000 & 14.560 & 0.009 & 2456784.3781 & 14.650 & 0.009 & 2456784.4144 & 14.454 & 0.015\\
2456784.3021 & 14.553 & 0.010 & 2456784.3802 & 14.624 & 0.010 & 2456784.4165 & 14.445 & 0.015\\
2456784.3043 & 14.597 & 0.010 & 2456784.3823 & 14.611 & 0.009 & 2456784.4187 & 14.453 & 0.015\\
2456784.3064 & 14.648 & 0.008 & 2456784.3845 & 14.568 & 0.008 & 2456784.4209 & 14.484 & 0.012\\
2456784.3085 & 14.725 & 0.008 & 2456784.3866 & 14.538 & 0.008 & 2456784.4230 & 14.475 & 0.012\\
2456784.3106 & 14.832 & 0.012 & 2456784.3887 & 14.515 & 0.006 & 2456784.4251 & 14.485 & 0.013\\
2456784.3128 & 14.922 & 0.010 & 2456784.3909 & 14.502 & 0.005 & 2456784.4272 & 14.492 & 0.011\\
2456784.3149 & 15.064 & 0.013 & 2456784.3930 & 14.521 & 0.003 & 2456784.4357 & 14.514 & 0.011\\
2456784.3170 & 15.166 & 0.017 & 2456784.3952 & 14.466 & 0.005 & 2456784.4379 & 14.591 & 0.007\\
2456784.3574 & 14.676 & 0.019 & 2456784.3973 & 14.477 & 0.006 & 2456784.4441 & 14.590 & 0.007\\
2456784.3595 & 14.684 & 0.016 & 2456784.3995 & 14.465 & 0.010 & 2456784.4461 & 14.579 & 0.010\\
2456784.3617 & 14.641 & 0.028 & 2456784.4016 & 14.472 & 0.010 & 2456784.4481 & 14.560 & 0.017\\
2456784.3678 & 14.662 & 0.028 & 2456784.4037 & 14.468 & 0.009 & 2456784.4501 & 14.586 & 0.020\\
2456784.3698 & 14.659 & 0.028 & 2456784.4059 & 14.474 & 0.009 & 2456784.4521 & 14.551 & 0.020\\
2456784.3719 & 14.636 & 0.029 & 2456784.4080 & 14.456 & 0.010  \\
\hline
   \end{tabular}
   \end{center}

\label{lastpage}

\end{document}